\documentclass[fleqn,twoside]{article}
\usepackage{espcrc2}

\usepackage{graphicx}
\usepackage[figuresright]{rotating}

\newcommand{\AmS}{{\protect\the\textfont2
  A\kern-.1667em\lower.5ex\hbox{M}\kern-.125emS}}

\title{GPCALMA: a Grid Approach to Mammographic Screening}

\author{
S.~Bagnasco\address[INFNTO]{I.N.F.N., Sezione di Torino, Italy.}, 
U.~Bottigli\address[SAS]{Struttura Dipartimentale di Matematica e Fisica dell'Universit\`a di Sassari and Sezione I.N.F.N. di Cagliari, Italy},
P.~Cerello\addressmark[INFNTO],
P.~Delogu\address[PISA]{Dipartimento di Fisica dell'Universit\`a and Sezione I.N.F.N. di Pisa, Italy},
M.~E.~Fantacci\addressmark[PISA],
E.~Lopez~Torres\address[CUBA]{CEADEN, Habana, Cuba},
G.~L.~Masala\addressmark[SAS],
P.~Oliva\addressmark[SAS],
A.~Retico\addressmark[PISA],
S.~Stumbo\addressmark[SAS]}
       
\begin{document}

\begin{abstract}
The next generation of High Energy Physics experiments requires a GRID approach to a 
distributed computing system and the associated data management: the key 
concept is the \"Virtual Organisation\" (VO), a group of geographycally distributed users 
with a common goal and the will to share their resources. A similar approach is 
being applied to a group of Hospitals which joined the GPCALMA project (Grid 
Platform for Computer Assisted Library for MAmmography), which will allow 
common screening programs for early diagnosis of breast and, in the future, 
lung cancer.
HEP techniques come into play in writing the application code, which makes use 
of neural networks for the image analysis and shows performances similar to 
radiologists in the diagnosis. GRID technologies will allow 
remote image analysis and interactive online diagnosis, with a relevant 
reduction of the delays presently associated to screening programs.
%The architecture adopted by GPCALMA will avoid data replication for all the 
%images with a negative diagnosis (about 95\% of the sample) and it will allow a 
%real time diagnosis for the 5\% of images with high cancer probability.
\vspace{1pc}
\end{abstract}

\maketitle

\section{Introduction}
A reduction of breast cancer mortality in asymptomatic women is possible in
case of early diagnosis \cite{1}, which is available thanks to screening programs, a periodical mammographic examination performed for $49$-$69$ years old women.
The GPCALMA Collaboration aims at the development of tools that would help 
in the early diagnosis of breast cancer: Computer Assisted Detection (CAD) would 
significantly improve the prospects for mammographic screening, 
by quickly providing reliable information to the radiologists. 

A dedicated software to search for massive lesions and microcalcification clusters was developed recently ($1998$-$2001$): its best results in the search for massive lesions (microcalcification clusters) are $94\%$ ($92\%$) for sensitivity and $95\%$ ($92\%$) for specificity.
Meanwhile, in view of the huge 
distributed computing effort required by the CERN/LHC collaborations, several 
GRID projects were started. 
It was soon understood that the application of GRID technologies to a database 
of mammographic images would facilitate a large-scale screening program, providing transparent and real time access to the full data set.

The data collection in a mammographic screening program will 
intrinsically create a distributed database, involving several sites with 
different functionality: data 
collection sites and diagnostic sites, i.e. access points from where 
radiologists would be able to query/analyze the whole distributed database.
The scale is pretty similar to that of LHC projects: taking Italy as an example, a full mammographic screening program would act on a target sample of about $6.8$ million women, thus generating
$3.4$ millions mammographic exams/year. 
With an average size of $60~MB$/exam, the amount
of raw data would be in the order of $200~TB$/year: a screening program on the European scale would be a data source 
comparable to one of the LHC experiments.

GPCALMA was proposed in 2001, with the purpose of developing a \"GRID 
application\", based on technologies similar to those adopted by the 
CERN/ALICE Collaboration. 
In each hospital, digital images will be stored in the local 
database and registered to a common service (Data Catalogue).
%managed by {\it AliEn}\cite{alien}. 
Data describing the mammograms, also 
known as metadata, will also be stored in the Data Catalogue and could be used to 
define an input sample for any kind of epidemiology study.
%Making use of the 
%ROOT and PROOF tools \cite{root}, 
The algorithm for the image analysis will be sent to the 
remote site where images are stored, rather than moving them to the 
radiologist's sites.
A preliminary selection of cancer candidates will be quickly 
performed and only mammograms with cancer probabilities higher than a given 
threshold would be transferred to the diagnostic 
sites and interactively analysed by one or more radiologists.
	
%Presently, a working version of the GPCALMA application is already available for local analysis. In parallel, a cluster of several PCs was already configured and successfully tested for the remote analysis of a set of 
%mammograms. 
%As soon as an {\it AliEn} managed database will be available (a dedicated {\it AliEn} Server was configured), the GPCALMA GRID-application will be able to dynamically select the input, making use of the {\it AliEn-PROOF} interface, whose first prototype was recently made available. 

\section{The GPCALMA CAD Station}
\begin{figure}[htb]
%\framebox[155mm]{\rule[-21mm]{0mm}{43mm}}
\centering
\includegraphics[width=77mm]{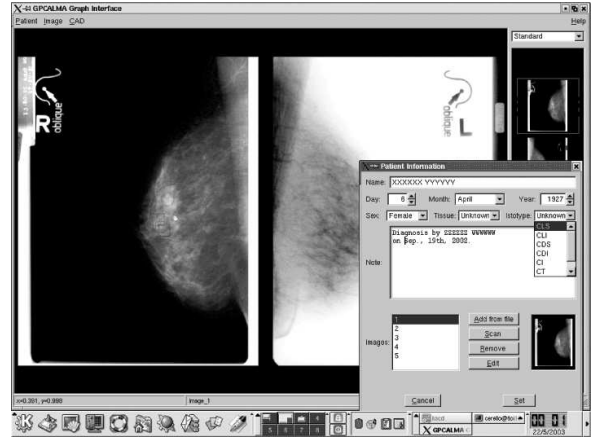}
\vspace{-1cm}
\caption{The GPCALMA Graphic User Interface. Three menus, corresponding to the Patient, the Images and the CAD diagnosis levels, drive it. On the left, the CAD results for microcalcifications and masses are shown in red squares and green circles, together with the radiologist's diagnosis (blue circle). On the right, the image colours are inverted. The widget drives the update of patient and image related metadata.} 
\label{GUI}
\end{figure}

The hardware requirements for the GPCALMA CAD Station are very simple: 
a PC with SCSI bus connected to a planar scanner and to 
a high resolution monitor. The station 
can process mammograms directly acquired by the scanner and/or images from 
file and allows human and/or automatic analysis of the digital mammogram. 
The software configuration for the use in local mode requires the installation 
of ROOT\cite{root} and GPCALMA, which can be downloaded either in the form of source code 
from the respective CVS servers.
The functionality is usually accessed through a Graphic User Interface, or,
for developers, the ROOT interactive shell.
The Graphic User Interface (fig. \ref{GUI}) allows the acquisition of new data, as well as the
analysis of existing ones. Three main menus drive the
creation of (access to) datasets at the patient and the image level and the
execution of CADe algorithms. 
The images are displayed according to the standard format required by 
radiologists: for each image, it is possible to 
insert or modify diagnosis and annotations, manually select the Regions of Interest (ROI) 
corresponding to the radiologists geometrical indication. An 
interactive procedure allows zooming, either continously or on a selected 
region, windowing, gray levels and contrast 
selection, image inversion, luminosity tuning. 
\vspace{-1cm}
The human analysis produces a diagnosis of the breast lesions in terms of 
kind, localization on the image, average dimensions and, if present, 
histological type. 
The automatic procedure finds the ROI's 
on the image with a probability of containing an interesting area larger than a pre-selected threshold value.

\section{Grid Approach}
The amount of data generated by a national 
or european screening program is so large that they can't be managed by a
single computing centre. In addition, data are generated according to an 
instrinsically distributed pattern: any hospital participating to the program 
will collect a small fraction of the data. Still, that amount would be large 
enough to saturate the available network connections.

\begin{figure}[htb]
\centering
\includegraphics[width=77mm]{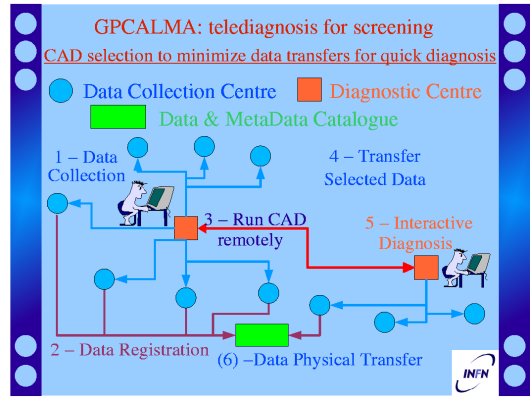}
\vspace{-1cm}
\caption{The screening use case: Data Collection Centres store and
register the images and the associated metadata in the {\it AliEn} Data Catalogue. Radiologists, from Diagnosis Centres, start the CAD remotely, without raw data transfer, making use of {\it PROOF}. Only the images corresponding to cancer probability larger than the selected threshold are moved to the Diagnosis Centre for the real-time visual inspection. Eventually, the small fraction of undefined cases can be sent to other radiologists.} 
\label{gpc22}
\end{figure}

The availability of the whole database to a radiologist, regardless of the data distribution, would provide several advantages:
\begin{itemize}
\item the CAD algorithms could be trained on a much larger data sample, with an
improvement on their performance, in terms of both sensitivity and specificity.
\item the CAD algorithms could be used as real time selectors of images with 
high breast cancer probability (see fig. \ref{gpc22}): radiologists would be able to prioritise their work, with a remarkable reduction of the delay between the data acquisition and the human diagnosis (it could be reduced to a few days).
\item data associated to the images (i.e., metadata) and stored on the 
distributed system would be available to select the proper input for epidemiology studies or for the training of young radiologists.
\end{itemize}

These advantages would be granted by a GRID approach: the configuration of
a Virtual Organisation, with common services (Data and Metadata Catalogue, Job Scheduler, Information System) and a number of distributed nodes providing 
computing and storage resources would allow the implementation of the screening, tele-training and epidemiology use cases.
However, with respect to the model applied to High Energy Physics, there are some important differences: the network conditions do not allow the transfer of large amounts of data, the local nodes (hospitals) do not agree on the raw data 
transfer to other nodes as a standard and, most important, some of the use cases require interactivity.

According to these restrictions, our approach to the implementation of 
the GPCALMA Grid application is based on two different tools: 
{\it AliEn} \cite{alien} for the management of common services, {\it PROOF} \cite
{root} for the interactive analysis of remote data without data transfer.
%\vspace{-1cm}

\subsection{Data Management}
The GPCALMA data model foresees several Data Collection Centres \ref{gpc22} , where mammograms are collected, locally stored and registered in the Data Catalogue. 
In order to make them available to a radiologist connecting from a Diagnostic
Centre, it is mandatory to use a mechanism that identifies the data 
corresponding to the exam in a site-independent way: they must be selected 
by means of a set of requirements on the attached metadata and identified 
through a Logical Name which must be independent of their physical location.
{\it AliEn} implements these features in its Data Catalogue Services, run by 
the Server: data are registered making use of a hierarchical namespace for their
Logical Names and the system keeps track of their association to the actual name of the 
physical files. In addition, it is possible to attach metadata to each level 
of the hierarchical namespace.
The Data Catalogue is browsable from the {\it AliEn} command line as well as 
from the Web portal; the C++ {\it Application Program Interface (API)} to ROOT is under development.
Metadata associated to the images can be classified in several categories:
patient and exam identification data, results of the CAD algorithm analysis, 
radiologist's diagnosis, histological diagnosis. etc..
Some of these data will be directly stored in the Data Catalogue, but some of them may be stored in dedicated files and registered: the decision will be made 
after a discussion with the radiologists.

\begin{figure}[htb]
%\framebox[155mm]{\rule[-21mm]{0mm}{43mm}}
\centering
\includegraphics[width=77mm]{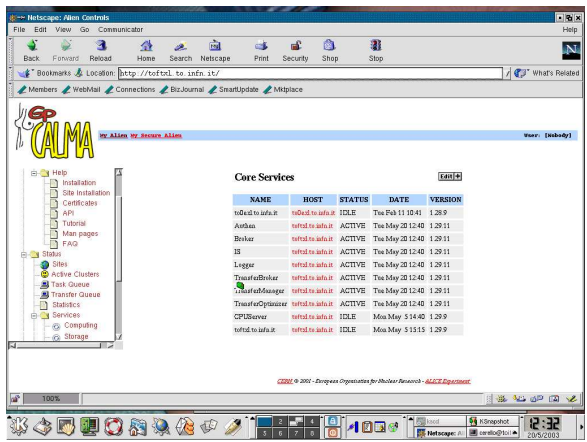}
\vspace{-1cm}
\caption{Screenshot from the GPCALMA AliEn WEB Portal. Making use of the left side frame, the site can be navigated. General Information about the AliEn project, the installation and configuration guides, the status of the Virtual 
Organisation Services can be accessed. On the main frame, the list of the 
core services is shown, together with their status.} 
\label{ga2}
\end{figure}

A dedicated {\it AliEn} Server for GPCALMA has been configured \cite{gpcalma}, 
in collaboration with the {\it AliEn} development team. Fig. \ref{ga2} shows a 
screenshot from the WEB Portal.
\vspace{-1cm}

\subsection{Remote Data Processing}
Tele-diagnosis and tele-training require interactivity in order to be fully 
exploited, while in the case of screening it would be possible - altough not optimal - to live without.
The PROOF {\it Parallel ROOt Facility} system \cite{root} allows to run interactive parallel processes on a distributed cluster of computers. 
A dedicated cluster of several PCs was configured and the remote analysis of a digitised mammogram without data transfer was recently run. 
As soon as input selection from the {\it AliEn} Data Catalogue will be possible, more complex use cases will be deployed. The basic idea is that, whenever a list of input Logical Names will be selected, that will be 
split into a number of sub-lists containing all the files stored in a given site and each sub-list will be sent to the corresponding node,
where the mammograms will be analysed.

\section{Present Status and Plans}
The project is developing according to the original schedule.
The CAD algorithms were rewritten in C++, making use of ROOT, in order to be 
 PROOF-compliant; moreover, the ROOT functionality allowed a significant improvement of the Graphic User Interface, which, thanks to the possibility to 
manipulate the image and the associated description data, is now considered 
fully satisfactory by the radiologists involved in the project.
The GPCALMA application code is available via CVS server for download and 
installation; a script to be used for the node configuration is being developed.The {\it AliEn} Server, which describes the Virtual Organisation and manages its services, is installed and configured; some {\it AliEn} Clients are in use, and they will soon be tested with GPCALMA jobs.
The remote analysis of mammograms was successfully accomplished making use of PROOF.
Presently, all but one the building blocks required to implement the tele-diagnosis and screening use cases were deployed. 
As soon as the 
implementation of the data selection from the ROOT shell through the 
{\it AliEn} C++ {\it API} will be available, GPCALMA
nodes will be installed in the participating hospitals and connected to the 
{\it AliEn} Server, hosted by INFN.
Hopefully, that task will be completed by the end of 2004.

\section{Acknowledgments}
The authors wish to thank the {\it AliEn} development team for their support and
guidance in the installation and configuration of the GPCALMA server.


\end{document}